# Controlling exchange bias in Co-CoO$_x$ nanoparticles by oxygen content

Miroslavna Kovylina[1], Montserrat García del Muro[1], Zorica Konstantinović[1], Manuel Varela[2], Òscar Iglesias[1], Amílcar Labarta[1], and Xavier Batlle[1]

[1]Departament de Física Fonamental and Institut de Nanociència i Nanotecnologia, Universitat de Barcelona, Martí i Franquès 1, 08028-Barcelona, Spain

[2]Departament de Física Aplicada i Òptica and Institut de Nanociència i Nanotecnologia, Universitat de Barcelona, Martí i Franquès 1, 08028-Barcelona, Spain

**Abstract.** We report on the occurrence of exchange bias on laser-ablated granular thin films composed of Co nanoparticles embedded in amorphous zirconia matrix. The deposition method allows controlling the degree of oxidation of the Co particles by tuning the oxygen pressure at the vacuum chamber (from $2\times10^{-5}$ to $10^{-1}$ mbar). The nature of the nanoparticles embedded in the nonmagnetic matrix is monitored from metallic, ferromagnetic (FM) Co to antiferromagnetic (AFM) CoO$_x$, with a FM/AFM intermediate regime for which the percentage of the AFM phase can be increased at the expense of the FM phase, leading to the occurrence of exchange bias in particles of about 2 nm in size. For oxygen pressure of about $10^{-3}$ mbar the ratio between the FM and AFM phases is optimum with an exchange bias field about 900 Oe at 1.8 K. The mutual exchange coupling between the AFM and FM is also at the origin of the induced exchange anisotropy on the FM leading to high irreversible hysteresis loops, and the blocking of the AFM clusters due to proximity to the FM phase.

**1. Introduction**

Magnetic nanoparticles [1] have attracted great interest during the last decades due to their technological applications in magnetic recording [2], magnetic resonance imaging [3,4], non-linear optics [5] and various biomedical applications [6]. As a result of their reduced dimensions, nanoparticles display magnetic and transport properties differing from bulk counterparts [1], as a consequence of the interplay among finite-size and surface effects, and interparticle interactions. Besides, particle surfaces may be easily oxidized resulting in core-shell structures, which can also be produced by physical and chemical procedures [7,8]. In particular, it is of interest the case when a ferromagnetic (FM) core is surrounded by an antiferromagnetic (AFM) shell, leading to the appearance of the so-called exchange bias (EB) phenomenon [9,10]. EB is a proximity effect [11] arising from the exchange coupling at the interface between a FM and an AFM which are in intimate contact. EB is usually described as an additional unidirectional anisotropy induced by the AFM material into the FM one, which yields a shift in the magnetic hysteresis loop along the magnetic-field axis, below the AFM ordering temperature. The magnitude of this shift is defined as the exchange bias field, H$_{eb}$. Although, EB was first observed in partially oxidized Co particles with a FM/AFM core/shell structure [9,12] most of the studies have focused on layered AFM/FM systems [13] due to their applications in advanced magnetic devices, such as read heads in magnetic recording and magnetic random access memories. Loop shifts as a function of the FM/AFM thickness ratio [10], coercive field enhancement [10], double hysteresis loops [11,14,15] and unusual magnetization reversal [16,17,18] are among the most investigated EB effects.

Nowadays, EB continues to attract a lot of interest from the fundamental and technological points of view, not only due to their potential applicability, but also because there is still a lack of microscopic understanding of the phenomena (for recent reviews see [10,16,19]). Moreover, it has been shown that control of the exchange interactions between the FM particle surface and an AFM embedding matrix can be a manner to beat the superparamagnetic limit [20,21]. It is worth noting that at the interface of core/shell nanoparticles there exist roughness and non-compensation of the magnetization which are two of the main ingredients in the models for EB in thin films [22,23]. Some of us recently showed that H$_{eb}$ in FM/AFM core/shell nanoparticles can be accounted for by a microscopic model taking into account the exchange interactions and net magnetization due to uncompensated spins both at the AFM/FM interface [24].

Some of the largest $H_{eb}$ have been reported for partially oxidized nanoparticles with FM/AFM core/shell structure, Co/CoO$_x$ being deserved special attention (loop shifts up to 10.2 kOe after field cooling the sample under 50 kOe have been reported for particles of 6 nm in size [25]; for a complete set of references see [16,19]). EB has also been studied in several nanoparticulate systems obtained by different physical fabrication techniques, e.g. gas-phase condensation method [26], ball milling and hydrogen reduction [27], sputtering [28,29], solid state reaction method [30], dual laser evaporation source [14] and pulsed laser ablation [31].

In this work, we report on the occurrence of EB in granular thin films composed of Co nanoparticles embedded in a Y-stabilized amorphous zirconia matrix, which have been deposited by laser ablation. This method allows controlling the degree of oxidation of the Co particles just by tuning the oxygen pressure at the vacuum chamber during deposition, without any posterior annealing. Consequently, we monitor the nature of the nanoparticles embedded in the nonmagnetic matrix from pure FM to pure AFM, with a FM/AFM intermediate regime for which the percentage of the AFM phase may be increased at the expense of the FM phase.

## 2. Experimental details

Granular films composed of Co nanoparticles embedded in amorphous zirconia ($ZrO_2$) matrix were obtained by pulsed laser ablation [32,33,34,35] in the presence of an oxygen partial pressure. Zirconia was stabilized with 7 mol% $Y_2O_3$ which provides the matrix with good oxidation resistance, thermal expansion coefficient matching that of metal alloys and very high fracture toughness values. Zirconia perfectly coats metallic nanoparticles and enables the occurrence of very sharp particle-matrix interfaces [32,33,34,35,36]. We used a KrF laser with wavelength of 248 nm, pulse duration of 34 ns and a composite rotary target to deposit simultaneously cobalt and zirconia. The pulse frequency (10 Hz), fluency (2 J/cm$^2$) and the laser spot (3 mm$^2$) were fixed for all samples, as well as the deposition temperature (295 K) and base pressure at the vacuum chamber (lower than $10^{-4}$ mbar). The Co/zirconia ratio in the composite target was also fixed for all samples and was chosen to grow granular films below the percolation threshold. We note that the sample grown at the base pressure had a volume Co fraction, $x_v$, of about 0.23. The degree of oxidation of the Co nanoparticles was controlled by varying the $O_2$ pressure in the chamber from $2\times10^{-5}$ (oxygen base pressure) to 0.1 mbar. It was assumed that the oxidation of the nanoparticles takes place essentially during their flight from the target to the substrate surface. Six samples were synthesized for the magnetic study at the following values of the $O_2$ pressure: $2\times10^{-5}$, $2.5\times10^{-4}$, $0.7\times10^{-3}$, $10^{-3}$, $10^{-2}$ and $10^{-1}$ mbar.

Average sample composition was determined by microprobe analyses. The films were structurally characterized by high resolution transmission electron microscopy (HRTEM) which was carried out on samples deposited onto a silicon nitride membrane window, allowing direct observation of as deposited samples. The particle size distributions were also obtained by fitting the high-temperature isothermal magnetization curves to a distribution of Langevin functions which models the superparamagnetic behaviour of the samples.

The degree of oxidation of the Co particles as a function of the oxygen partial pressure was analyzed by X-ray photoelectron spectroscopy (XPS). The XPS spectra for the O 1s, Co $2p_{3/2}$ and $2p_{1/2}$, Zr $3d_{5/2}$ and $3d_{3/2}$, Y $3d_{5/2}$ and $3d_{3/2}$ and C 1s core levels were recorded using the Al K$_\alpha$ emission line (h$\nu$=1486.6 eV; incident angle of the beam=45°). Spectra were obtained before and after 10 minutes of in situ low-energy (4 keV) sputtering process (incident at 45°) to avoid the contribution from the oxidized particles at the free surface of the thin films. Energy calibration was carried out by adopting the C 1s core level, associated with the usual surface contamination layer (binding energy $E_B$=284.8 eV), as reference peak [37].

Magnetization measurements were carried out in a SQUID magnetometer. The temperature dependence of the magnetization under 50 Oe after field cooling (FC) and zero-field cooling (ZFC) the sample, and the hysteresis loops at various temperatures up to a maximum field of 50 kOe were recorded.

**3. Structural characterization**

Average sample composition was determined by electron microprobe analyses being $x_v$ (the Co volume fraction) below 0.23 and decreasing with increasing oxygen pressure. For all the studied samples the Co content was well below the percolation threshold $x_c \sim 0.35$ [34].

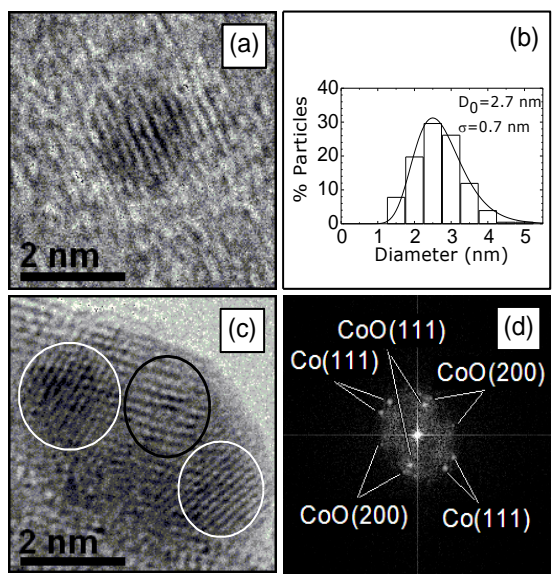

Figure 1. HRTEM images for the samples deposited at $P_{O_2}$: (a) $2\times10^{-5}$ and (c) $5\times10^{-4}$ mbar. White lines in (c) circle Co regions while black line circles $CoO_x$ ones. The size distribution determined from HRTEM images corresponding to the sample deposited at the base pressure of $2\times10^{-5}$ mbar is shown in panel (b). Panel (d) shows the Fourier transform of the image in figure 2(c), where the obtained diffraction pattern was indexed to both fcc Co and CoO reflections.

Figure 1 shows HRTEM images corresponding to a sample with metallic Co nanoparticles, obtained at the base oxygen pressure in the vacuum chamber, and another one grown at $P_{O_2}=5\times10^{-4}$ mbar. As an example of the first case, figure 1(a) shows a bright field HRTEM of a nanoparticle where the dark regions correspond to Co and the light ones to the amorphous $ZrO_2$ matrix. The microstructure is very clean: the lattice fringes indicate single crystalline Co particles and the particle-matrix interfaces are sharp, as observed in many other granular systems prepared by pulse laser deposition [34,35]. Selected area electron diffraction patterns can be fitted to metallic fcc Co. Figure 1(c) shows the particle-size distribution obtained from HRTEM images corresponding to the sample prepared at the oxygen base pressure. Average particle diameter is about 2.7 nm with a standard deviation of 0.7 nm, which indicates a narrow size distribution. For the second case (see figure 1(c) for the sample grown at $P_{O_2}=5\times10^{-4}$ mbar), several lattice fringes, oriented in different directions, can be observed within one particle together with an amorphous irregular halo surrounding the particle without any clear interface. Figure 1(d) shows the Fourier transform of the image in figure 1(c), where the diffraction pattern can be indexed to both Co and CoO reflections. The crystallographic interplanar distances, determined from the diffraction pattern in figure 1(d), are 0.200±0.007 nm for Co(111) (expected value: 0.2046 nm), 0.25±0.01 nm for CoO(111) (expected value for CoO: 0.2455 nm) and 0.22±0.01 for CoO(200) (expected value for CoO: 0.2126 nm), in good agreement with the expected values for both metallic and oxidized cobalt. Therefore, within these polycrystalline nanoparticles some areas correspond to Co regions while others correspond to $CoO_x$. Besides, particles seem to be larger in the oxidized sample probably due to the fact that the presence of $O_2$ at the vacuum chamber during the deposition impedes the small particles arriving on the substrate and results, together with the cell expansion due to oxidation, in the shift of the particle size distribution to higher values. Amorphous $CoO_x$ may be very difficult to be distinguished from the zirconia matrix, so $CoO_x$ clusters are probably only observed when they are aggregated with metallic Co particles. We note that the zirconia matrix starts to crystallize when the sample is overexposed under high-electron current.

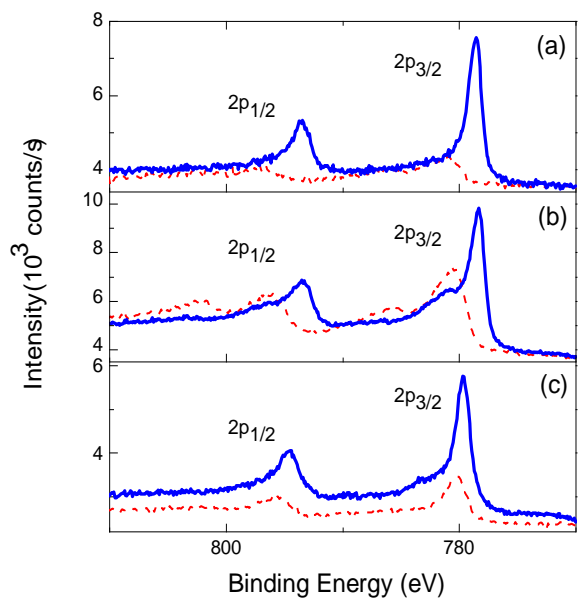

Figure 2. (colour online) Co $2p_{3/2}$ and $2p_{1/2}$ core levels from XPS data for pre-sputtering (red dashed line) and post-sputtering (blue solid line) spectra of the samples for $P_{O_2}$: (a) $2\times10^{-5}$, (b) $10^{-3}$, and (c) $10^{-1}$ mbar.

Further information about the degree of oxidation of the Co particles as a function of oxygen partial pressure can be gained by analyzing XPS data. Figure 2 shows the Co $2p_{3/2}$ and $2p_{1/2}$ core levels for three values of the oxygen partial pressure, both before and after sputtering the surface. As expected, the only peaks observable before sputtering correspond to Co-O bonds (at about 780.0-780.7 eV and 795.4-796.2 for the $2p_{3/2}$ and $2p_{1/2}$ contributions, respectively, in figure 2), while there is no trace of Co-Co bonds characteristic of metallic Co which should appear at 778.1-778.3 eV ($2p_{3/2}$) and 793.1-793.3 eV ($2p_{1/2}$). For a general reference on binding energies see for example [38]. In contrast, once the oxidized surface layer is removed by sputtering, the degree of oxidation of the remaining Co nanoparticles increases with increasing oxygen partial pressure in the chamber during deposition. For the sake of simplicity just three examples are shown being representative of the overall behavior. For the lowest oxygen pressure ($P_{O_2}=2\times10^{-5}$ mbar) only the peaks corresponding to metallic Co-Co bonds are observed (see figure 2(a)), in agreement with HRTEM image in figure 1(a). In the case for $P_{O_2}=10^{-3}$ mbar (figure 2(b)) the peaks corresponding to both Co-Co and Co-O bonds are observed indicating that the particles are partially oxidized, in agreement with HRTEM image in figure 1(c). At about the highest oxygen pressure in this work ($P_{O_2}=0.1$ mbar, figure 2(c)) Co nanoparticles are almost completely oxidized since the peaks corresponding to the Co-O bonds are the unique contribution to the 2p core level. The XPS data confirm that Co clusters get oxidized during their flight from the target to the substrate due to reaction with the remaining oxygen in the deposition chamber. The difference in the peak intensity between pre-sputtering and post-sputtering spectra in figure 2 arises from the distinct time of exposure (longer for the case of post-sputtering spectra). The peaks corresponding to Zr $3d_{5/2}$ and $3d_{3/2}$ core levels, which are observable before and after sputtering the surface of the samples, are associated with Zr-O bonds (at about the expected 182-182.5 eV and 184-184.5 eV for the $3d_{5/2}$ and $3d_{3/2}$ contributions, respectively [38]), without any indication of the existence of metallic Zr-Zr bonds. Besides, XPS spectra for Zr core levels do not show any significant dependence on the partial oxygen pressure in the chamber. As far as Co is concerned, the Co/CoO atomic fraction is about 0.9±0.1 for the sample grown at $P_{O_2}=10^{-3}$ mbar (figure 2(c)).

## 4. Magnetic characterization

Figure 3 shows the ZFC and FC curves of samples obtained at different $P_{O_2}$. Figure 3(a) shows the case corresponding to metallic, FM Co nanoparticles embedded in the zirconia matrix (ablated at the base pressure) which exhibits all the features that are characteristic of a narrow distribution of slightly interacting small FM particles. As $P_{O_2}$ increases, a variety of effects can be observed: (i) The magnetization of the samples decreases due to the decrease in the FM fraction, at the expense of the increase in the AFM $CoO_x$ phase (see below); (ii) The position of the ZFC peak shifts and broadens from 6.5 K for $P_{O_2}= 2\times10^{-5}$ mbar to 12.5 K for $P_{O_2}=10^{-3}$ mbar, suggesting the occurrence of exchange coupling between the AFM and FM phases [20,21], and/or the shift to higher values of the particle size distribution; (iii) The increase in the temperature of the onset of the ZFC-FC irreversibility suggests the existence of interacting FM clusters inside partially oxidized aggregates, as HRTEM images also support, which broaden the distribution of particle relaxation times; and (iv) The increase in the FM background at high temperatures which indicates the presence of some large FM aggregates.

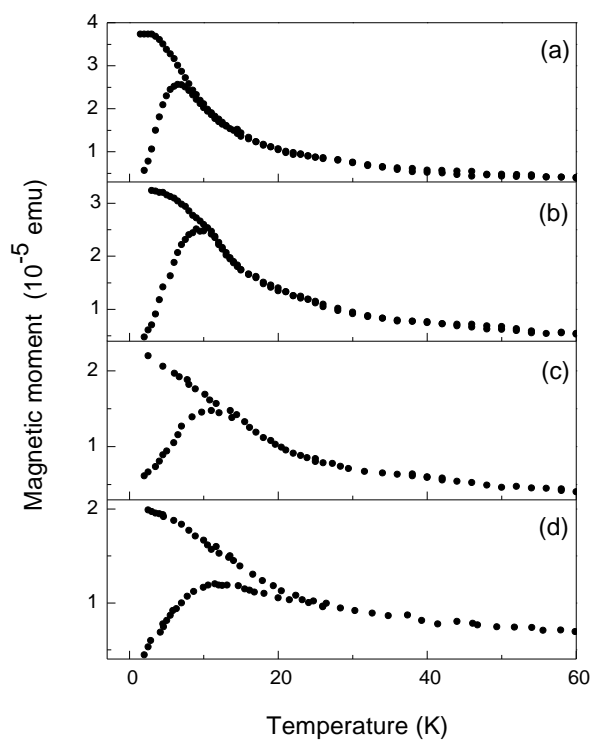

Figure 3. Temperature dependence of the field cooled (FC) and zero field cooled (ZFC) magnetizations measured at 50 Oe for the samples prepared at $P_{O_2}$: (a) $2\times10^{-5}$, (b) $2.5\times10^{-4}$, (c) $7\times10^{-4}$ and (d) $10^{-3}$ mbar, corresponding to Co volume fraction $x_v$: (a) 0.23, (b) 0.22, (c) 0.21, and (d) 0.19, as determined from electron microprobe analyses.

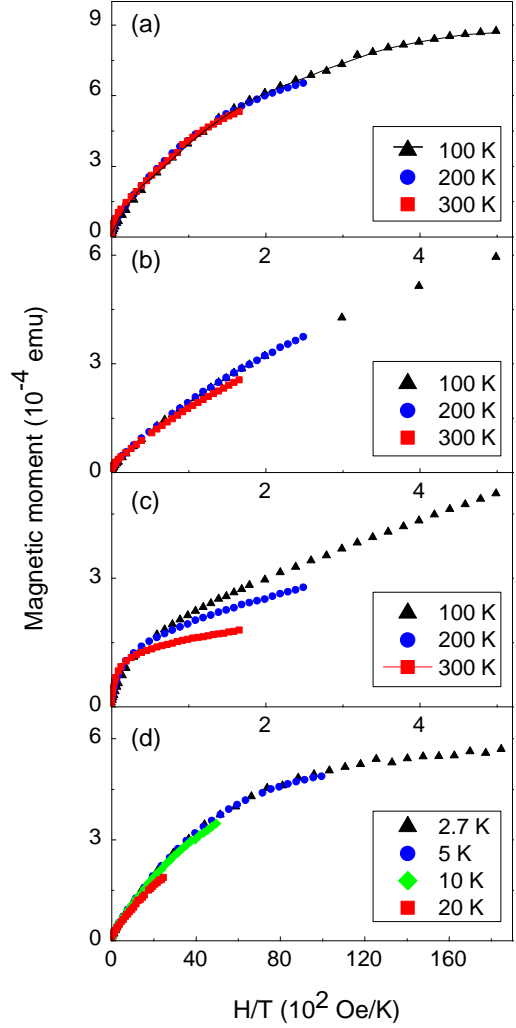

Figure 4. (colour online) Magnetization against H/T for the samples prepared at Po$_2$: (a) 2×10$^{-5}$, (b) 7×10$^{-4}$, (c) 10$^{-3}$, and (d) 0.1 mbar. For (a-c), the triangles (black) correspond to 100 K, the circles (blue) to 200 K, and the squares (red) to 300 K. For (d), the triangles (black) correspond to 2.7 K, the circles (blue) to 5 K, the diamonds (green) to 10 K and the squares (red) to 20 K.

One might expect that in the superparamagnetic regime magnetization curves should scale when plotted as a function of H/T. Figure 4 shows such plot for four representative cases with increasing Po$_2$ and, hence, AFM/FM ratio for several temperatures. The magnetization curves for metallic FM Co nanoparticles scale perfectly (figure 4(a)), while the quality of this scaling gets gradually worse as Po$_2$ increases (see figures 4(b) and 4(c)). The scaled curves for the sample ablated at the base pressure can be well fitted to a log-normal distribution of Langevin functions (solid line in figure 4(a)), leading to a mean particle size of 1.8 nm and standard deviation of 0.2 nm, in agreement with the ZFC curve in figure 3(a) and the size distribution in figure 1(b), suggesting a very narrow distribution of FM particles. In this fit, it was assumed that all particles have the Co bulk saturation magnetization of 0.15 µ$_B$/Å$^3$. Therefore, the slight discrepancy between the obtained values of the average crystalline and magnetic sizes could arise from a reduced value of the particle magnetization due to surface disorder and finite-size effects. The effect of the progressive oxidation of the FM particles with increasing oxygen pressure may be well understood in figure 4(c) where magnetization curves result from the superimposition of two superparamagnetic contributions: the magnetization of the FM components through Langevin terms and an AFM component which yields a linear term on the magnetic field. While the Langevin terms scale on H/T, this is not the case of the AFM components as reported in [39]. This is the main reason for the lack of scaling in figures 4(b) and 4(c). Assuming that the AFM components arise from superparamagnetic $CoO_x$ clusters, the magnetization curves in figures 4(b) and 4(c) at 300 K, which is close to the Néel temperature of bulk CoO, can be fitted to a distribution of Langevin functions without a linear AFM term, yielding a mean particle size of 2.7 nm with a standard deviation of 0.3 nm and 3.4 nm with a standard deviation of 0.4 nm, respectively. The increases in the mean particle size and standard deviation are in agreement with the shift and broadening of the ZFC curves in figure 3. It is worth noting that the ZFC curves measured at low fields are mostly due to the FM contribution while both AFM and FM components contribute to the magnetization curves as a function of H. Interestingly enough, figure 4(d) shows magnetization curves for a case with completely oxidized Co particles ($P_{O_2}$=0.1 mbar) within the low temperature range 2.7 to 20 K. As the FM component has disappeared completely, the H/T scaling resembles that of a distribution of superparamagnetic AFM particles with uncompensated spins [39]. The magnetization curves can be fitted to a distribution of Langevin functions yielding a mean value of the uncompensated spins per cluster below about 5 $\mu_B$, suggesting that the AFM clusters are very small.

The drastic modification of the magnetic properties of the granular films depending on the AFM/FM ratio can be further investigated in figure 5 where the hysteresis loops at 5 K after a ZFC process are depicted. Figure 5(a) shows the characteristic hysteresis loop of an assembly of blocked FM Co nanoparticles as expected for the sample deposited at the base pressure. With the increase in $P_{O_2}$ an additional AFM contribution superimposes the central FM hysteresis loop. Up to about $P_{O_2}=10^{-3}$ mbar there still survives a significant FM fraction which is blocked at 5 K and is exchange coupled to the AFM clusters, such that the AFM regions are also blocked by proximity to the FM (see figures 5(b) and 5(c)). This proximity effect and the resulting induced anisotropy are at the origin of the high irreversibility in the hysteresis loops in figures 5(b) and 5(c). By further increasing $P_{O_2}$ the amount of the AFM component grows at the expense of the FM one, such that as the exchange coupling progressively disappears, the AFM component becomes superparamagnetic and the irreversibility in the hysteresis loops vanishes (see figures 5(d) and 5(e)). All the foregoing demonstrates that it is required to have a minimum amount of both FM and AFM phases for exchange coupling to show up [16,20].

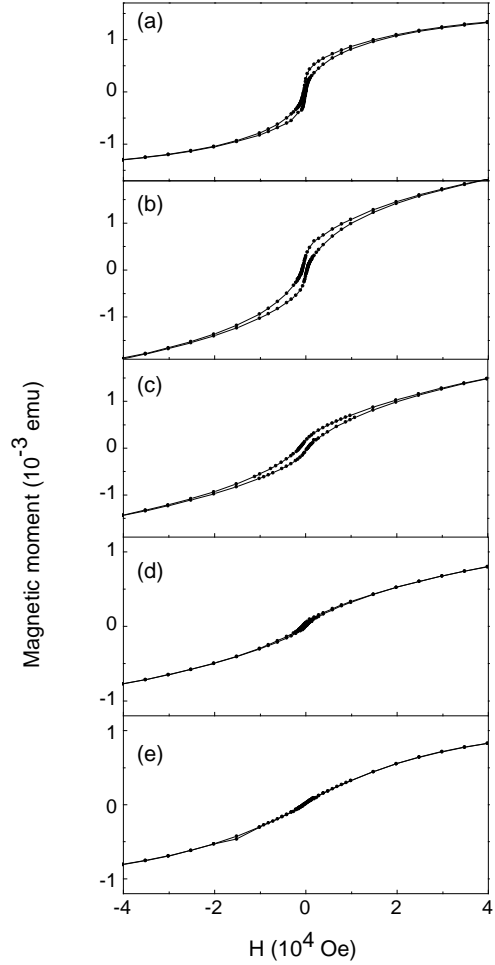

Figure 5. (colour online) Hysteresis loops at 5 K after a ZFC process for the samples prepared at $P_{O_2}$: (a) $2\times10^{-5}$, (b) $2.5\times10^{-4}$, (c) $7\times10^{-4}$, (d) $10^{-2}$ and (e) $10^{-1}$ mbar.

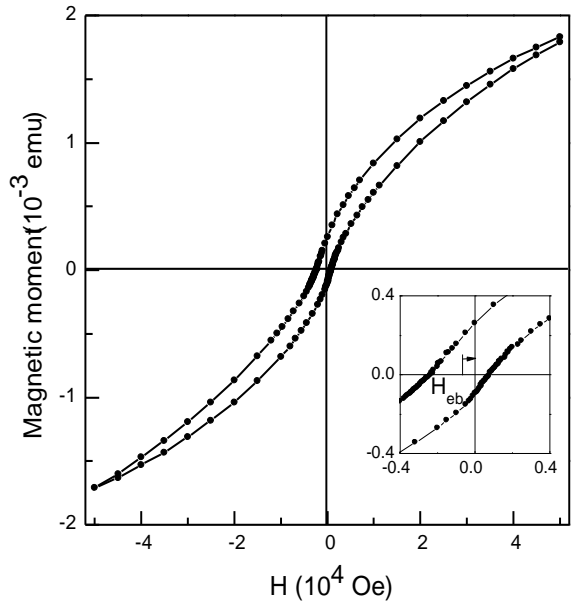

Figure 6. Hysteresis loop at 1.8 K for $P_{O_2}=10^{-3}$ mbar recorded after field cooling the sample under 50 kOe. The inset shows a detail of the low field region where the shift of the hysteresis loop due to the EB is clearly observable. $H_{eb}$ is indicated in the inset by a small arrow.

## 5. Exchange bias

In order to gain a deeper insight on the nature of the exchange coupling phenomenon taking place between the FM and AFM phases as a function of $P_{O_2}$, hysteresis loops after field cooling the sample at 50 kOe from room temperature down to the measuring temperature were recorded within the range 1.8-16 K. Figure 6 shows an example where a shifted loop is observed as a consequence of EB between the FM and AFM phases, showing $H_{eb}$ = 900 Oe at 1.8 K. Besides, the magnetic irreversibility is evident. Note that, in this case, the maximum applied field was lower than the irreversibility field, so the observed loop shift may not entirely correspond to an exchange bias phenomenon because the measured hysteresis loop could be a minor loop.

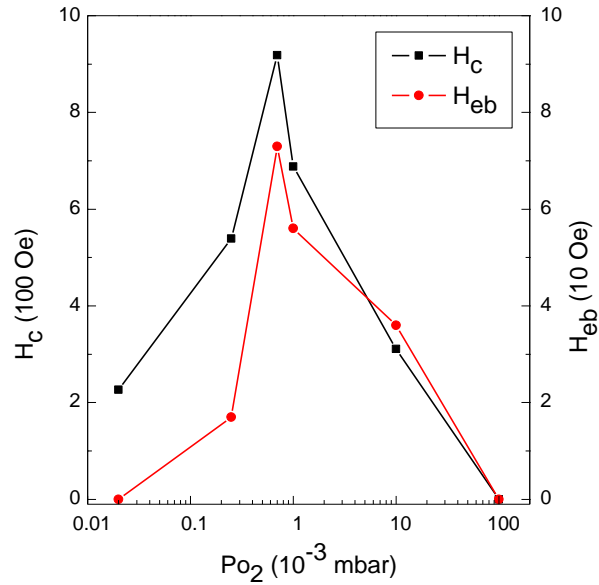

Figure 7. (colour online) Exchange bias ($H_{eb}$) and coercive field ($H_c$) dependence on oxygen pressure ($P_{O_2}$) at the deposition chamber.

The exchange bias ($H_{eb}$) and coercive field ($H_c$) as functions of $P_{O_2}$ are shown at 5 K in figure 7. Both $H_c$ and $H_{eb}$ progressively increase with the oxidation degree of the metallic component up to about $P_{O_2}=0.7\times10^{-3}$ mbar, yielding maximum values of $H_c^{max}$=918 Oe and $H_{eb}^{max}$ =73 Oe at 5 K. Consequently, the optimum ratio between the FM and AFM components takes place at about this pressure. XPS data (figure 2(b)) suggest that this optimum value corresponds roughly to half of the Co atoms being oxidized. The fact that the coercive field of the FM phase also increases as the EB develops gives further support to the existence of exchange coupling. Above that optimum pressure both $H_c$ and $H_{eb}$ decrease and vanish completely at about $P_{O_2}=0.1$ mbar as the FM phase disappears. This evolution is associated with the progressive change in the FM/AFM ratio from metallic FM Co - characterized by small $H_c$ and zero $H_{eb}$ - to the pure AFM $CoO_x$ with zero $H_c$ and $H_{eb}$. For the intermediate cases the characteristic exchange bias behaviour arising from the exchange coupling between the FM and AFM phases is observed.

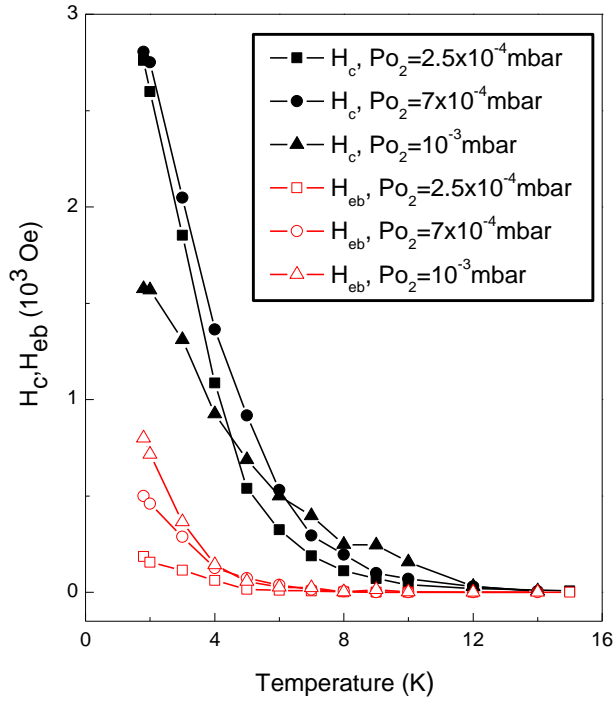

Figure 8. (colour online) Temperature dependence of the exchange bias ($H_{eb}$) (empty red symbols) and coercive field ($H_c$) (solid black symbols) for the samples prepared at $Po_2$: $2.5\times10^{-4}$ (squares), $7\times10^{-4}$ (circles), and $10^{-3}$ (triangles) mbar.

The dependences of $H_c$ and $H_{eb}$ on temperature are shown in figure 8. A remarkable maximum value of $H_{eb}$=900 Oe is observed for $Po_2=10^{-3}$ mbar at 1.8 K. The decrease of $H_c$ and $H_{eb}$ with increasing temperature follows the onset of superparamagnetism as displayed in ZFC-FC curves. $H_{eb}$ increases gradually for all temperatures with the increase of $Po_2$ during the ablation from $2.5\times10^{-4}$ to $10^{-3}$ mbar. $H_c$ follows a more complex trend. At 1.8 K, $H_c$ for $Po_2=10^{-3}$ mbar is less than that for $7\times10^{-4}$ mbar due to the magnetic frustration associated with the FM-AFM interactions as typically observed in disordered magnets. This magnetic frustration also produces a smoother $H_c(T)$ for $Po_2=10^{-3}$ mbar than those for the other two samples in figure 8, for which $H_{eb}$ is smaller, yielding a crossing among the $H_c(T)$ curves. This effect results in the stabilization of the FM phase due to the exchange coupling to the AFM phase.

**6. Conclusions**

The occurrence of exchange bias has been shown in partially oxidized Co particles of about 2 nm. This size is in the lowest limit reported for the occurrence of EB in Co/CoO core/shell structures (3 nm) [40] and Co clusters embedded in a CoO matrix (2.5 nm) [41]. This critical size might be due to the exchange energy at the FM/AFM interface becoming smaller than both the effective Zeeman energy of the FM and the anisotropy energy of the AFM. For oxygen pressure of about $10^{-3}$ mbar the ratio between the FM Co and AFM $CoO_x$ phases is optimum (about 50% each) with an exchange bias field close to 1 kOe at 1.8 K after a FC of 50 kOe. The occurrence of this exchange bias field may be related to the polycrystalline nature of the nanoparticles in which Co and $CoO_x$ nanocrystalline clusters coexist in intimate contact. The mutual exchange coupling between the AFM and FM phases due to proximity results not only in loop shifts but also in: i) the blocking of small AFM clusters when the FM phase is also blocked; ii) the high irreversibility in the hysteresis loops due to the induced exchange anisotropy and magnetic frustration associated with AFM-FM interactions as observed in many magnetically disordered systems; and iii) the thermal stabilization of the FM as both observed in the smooth temperature dependence of the coercive field and the increase in the temperature of the maximum of the ZFC curve.

**Acknowledgements**

We would like to thank the staff of the Scientific and Technical Facilities of the University of Barcelona. Financial support of the Spanish CICYT (MAT2006-03999) and Catalan DURSI (2005SGR00969) are gratefully recognized. MK thanks the Spanish MEC for the financial support through a PhD grant.